# Semiconducting character of LaN: magnitude of the band gap, and origin of the electrical conductivity

Zihao Deng[1] and Emmanouil Kioupakis[1,*]

[1]Department of Materials Science and Engineering, University of Michigan, Ann Arbor, Michigan, 48109, United States

*Correspondence to: kioup@umich.edu

Lanthanum nitride (LaN) has attracted research interest in catalysis due to its ability to activate the triple bonds of $N_2$ molecules, enabling efficient and cost-effective synthesis of ammonia from $N_2$ gas. While exciting progress has been made to use LaN in functional applications, the electronic character of LaN (metallic, semi-metallic, or semiconducting) and magnitude of its band gap have so far not been conclusively determined. Here, we investigate the electronic properties of LaN with hybrid density functional theory calculations. In contrast to previous claims that LaN is semi-metallic, our calculations show that LaN is a direct-band-gap semiconductor with a band-gap value of 0.62 eV at the X point of the Brillouin zone. The dispersive character of the bands near the band edges leads to light electron and hole effective masses, making LaN promising for electronic and optoelectronic applications. Our calculations also reveal that nitrogen vacancies and substitutional oxygen atoms are two unintentional shallow donors with low formation energies that can explain the origin of the previously reported electrical conductivity. Our calculations clarify the semiconducting nature of LaN and reveal candidate unintentional point defects that are likely responsible for its measured electrical conductivity.

Nitride compounds are a rich class of functional materials. The main-group III-nitrides are important semiconductors that find applications in electronics, optoelectronics, and photocatalysis. Recently, transition-metal and rare-earth nitrides have attracted attention due to their promise in, e.g., piezoelectric,[1] superconducting,[2] and catalysis applications.[3,4] In particular, Ye *et al.* discovered that lanthanum nitride (LaN) facilitates stable and highly efficient ammonia synthesis through the activation of $N_2$ gas by nitrogen vacancies at the surface. LaN shows a catalytic performance that is comparable to ruthenium-based catalysts but at a much lower cost.[4] Thus, LaN is a promising nitride material for catalysis and other chemical applications.

While LaN exhibits great promise in functional applications, one fundamental question that has not been fully addressed is whether LaN has metallic, semi-metallic, or semiconducting nature. Understanding this fundamental electronic character of LaN is critical for its future applications. Many previous theoretical calculations have attempted to elucidate the electronic band structure of LaN. Early density functional theory (DFT) calculations with the augmented plane wave method (APW) showed that the conduction and valence band of LaN overlap by up to 40 mRy, indicating a semi-metallic nature.[5,6] Vaitheeswaran *et al* studied the electronic properties of LaN using tight-binding linear muffin-tin orbitals with the local-density approximation (LDA) to the exchange-correlation functional and found a metallic nature for LaN. They also estimated the superconducting transition temperature to be 0.65 K.[7] Later calculations with the generalized gradient approximation (GGA) functional observed the overlap between valence and conduction band in the band structure and characterized LaN either as metallic or semi-metallic.[8,9] However, these results contradict to the calculations using the hybrid screened-exchange local density approximation (sX-LDA) functional where an indirect band gap of 0.75 eV was found, suggesting that LaN might be a semiconductor.[10] Recently, more calculations seem to support the semiconducting nature of LaN. Gupta *et al* used LDA functional and found an indirect band gap of 0.5 eV.[11] GGA+$U^{SIC}$ functional was employed by Meenaatci *et al* and an indirect band gap of 0.65 eV was found.[12] However, by using the LSDA+U functional, Larson *et al* discovered a small direct band gap of 0.4 eV for LaN.[13] Similarly, a direct band gap of 0.6 eV was recently obtained by MBJLDA functional for both wurtzite and rocksalt LaN.[14,15] In addition, Sreeparvathy *et al* found a direct band gap of 0.814 eV using full potential linearized augmented plane wave (FP-LAPW) method with TB-mBJ functional,[16] which seems to agree with the experimental band gap of 0.82 eV measured from optical absorption.[17] Despite the progress from previous theoretical studies, the nature (direct or indirect) and the magnitude of the band gap are not conclusively determined for LaN due to different methods and functionals employed in the calculations, which necessitates a re-investigation of the electronic properties of LaN with modern electronic structure calculations.

Since LaN is found by the more advanced functionals to exhibit a band gap, and thus to present a semiconducting rather than a semi-metallic character, its electrical conductivity must originate from intrinsic or unintentional dopants. However, to the best of our knowledge, there is no theoretical investigation into the thermodynamics of the intrinsic defects and common impurities of LaN, which is the key to understand the origin of its electrical conductivity. This knowledge is also necessary in order to rationally tune its conductivity by controlling the defect formation and doping in experiments. Thus, theoretical insights into the intrinsic defect formation and ionization energies are crucial to enable the adoption of LaN in wider electronic and catalysis applications.

In this work, we study the electronic properties (band structure, effective masses, dielectric constants, etc.) and defect thermodynamics of LaN using first-principles calculations based on DFT with the HSE06 hybrid functional,[18,19] which predicts accurate electronic properties for a wide range of materials. We find that LaN is a direct-band-gap semiconductor with a band gap of 0.62 eV at the X point of the Brillouin zone. Our defect calculations attribute the origin of its electrical conductivity to the unintentional formation of N vacancies or substitutional O impurities. Our calculations clarify the semiconducting nature of LaN and reveal candidate defects that are the likely origin of its measured electrical conductivity.

DFT calculations were performed using the Vienna ab-initio Simulation Package (VASP).[20] GW-compatible Perdew-Burke-Ernzerhof (PBE) pseudopotentials[21] for La and N and a plane-wave energy cutoff of 500 eV were employed in all calculations. The La $5s^2 5p^6 5d^1 6s^2$ and N $2s^2 2p^3$ were treated as valence electrons. In order to get accurate electronic properties, we used the HSE06 hybrid exchange-correlation functional with a standard mixing parameter of 0.25.[18,19] The electronic band structures of LaN were calculated with a fully relaxed 2-atom rocksalt primitive cell (Space group *Fm-3m*) and a Γ-center 8×8×8 Brillouin zone sampling grid.[22] The special k-point path for plotting the band structure followed the convention of Setyawan and Curtarolo.[23] Spin-orbit coupling effects were included in the band structure. The static and high-frequency dielectric constants were calculated by the self-consistent response to finite electric field at the HSE level using modern theory of polarization.[24–27] Electron and hole effective masses were extracted by fitting the HSE band structure with the hyperbolic equation,

$$E(k) = \frac{\mp 1 \pm \sqrt{1+4\alpha \frac{\hbar^2 k^2}{2m^*}}}{2\alpha} + E_0 \qquad (1)$$

where $E_0$ is the energy of the band extremum, $m^*$ is the effective mass, and $\alpha$ is a fitting parameter to characterize the non-parabolicity of the band. The band alignment of LaN was obtained by aligning the bulk average electrostatic potential to the vacuum level. This was done by performing HSE calculations for the LaN (100) and (110) slabs without surface relaxation. The alignment results for the two slabs differ by only 70 meV. Defect calculations[28] were performed with 2×2×2 supercells built from the 8-atom rocksalt unit cell (i.e., 64 atoms) and the Brillouin zone was sampled with a Γ-center 2×2×2 grid. The defect formation energies and charge-transition levels for the $O_N$ defect are found to change by less than 0.05 eV for a larger supercell with 128 atoms (Table S1). Thus, the 64-atom supercell was employed for all subsequent defect studies to reduce the computational cost. All defect supercells were relaxed with HSE06 by allowing ion displacements until the forces on the ions were less than 0.02 eV Å$^{-1}$. Spin polarization was included in calculations with unpaired electrons. We employed the scheme of Freysoldt *et al.*[29] and our calculated static dielectric constant for LaN to correct the artificial periodic charged-defect interactions. The competing phases we considered in the thermodynamic analysis of defect formation are $La_2O_3$, $NO_2$, $NH_3$, and all stable elemental phases.

We first investigate the band structure of LaN to determine its electronic character (metal or semiconductor) and the magnitude of its band gap, if one exists. Previous DFT calculations using the GGA functional characterized LaN as a semimetal.[8,9] Indeed, our own PBE band-structure calculations (Figure S1) reveal that the conduction and valence bands overlap by 115 meV at the X point, which leads to the conclusion that LaN is a semimetal and agrees with many previous theoretical studies. However, calculations with semilocal exchange-correlation functionals such as LDA or PBE severely underestimate the band gaps of materials and may erroneously lead to a closed gap in LaN. Moreover, we find that even after one-shot GW corrections on top of PBE, the band gap is still extremely small (0.05 eV from Table 1), which indicates that the metallic PBE state is a poor starting point for GW. Thus, the adoption of non-local functional in the calculations is necessary to give accurate band gap and a good starting point for GW corrections. Table 1 shows our calculated band gap for LaN using the HSE06 hybrid functional with GW and spin-orbit coupling corrections, respectively. Under all circumstances, LaN shows a direct band gap at the X point of the Brillouin zone. The band gap value is 0.75 eV if we only use HSE in the calculations. Both one-shot GW correction and spin-orbit coupling decrease the band gap value into the range of 0.6-0.7 eV. Our results are consistent with the most recent studies by Winiarski and Kowalska where they found a direct band gap of 0.6 eV for LaN using MBJLDA functional including the relativistic spin-orbit effect.[14,15] This provide a direct evidence that LaN is a semiconductor rather than a semimetal. We therefore attribute the miscategorization of LaN as a semimetal in previous studies to the systematic band-gap underestimation problem of LDA and GGA.

TABLE 1. Comparison of the LaN band gap calculated in this work and previous studies

| Method | $E_{gap}$ (eV) | Character |
|---|---|---|
| PBE (this work) | 0 | semi-metallic |
| PBE+GW (this work) | 0.05 | direct X-X |
| HSE06 (this work) | 0.75 | direct X-X |
| HSE06+GW (this work) | 0.68 | direct X-X |
| HSE06+SOC (this work) | 0.62 | direct X-X |
| APW[5,6], LDA[7], GGA[8,9], HGH[11] | 0 | semi-metallic |
| sx-LDA[10] | 0.75 | indirect Γ-X |
| LDA[11] | 0.5 | indirect Γ-X |
| GGA+U[12] | 0.65 | indirect Γ-X |
| LSDA+U[13] | 0.4 | direct X-X |
| MBJLDA[14,15] | 0.6 | direct X-X |
| FP-LAPW, TB-mBJ[16] | 0.814 | direct X-X |

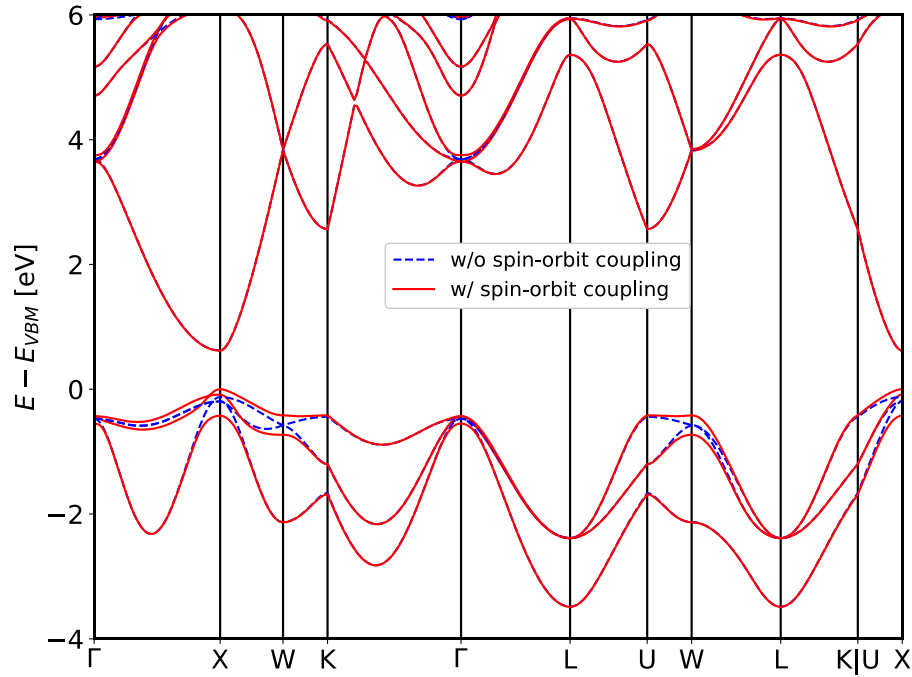

FIG. 1. The electronic band structure of LaN calculated with the HSE06 hybrid density functional with (w/) and without (w/o) spin-orbit coupling. Instead of a semimetal, LaN is a direct band gap semiconductor with a band gap of 0.62 eV at the X point.

Figure 1 shows the electronic band structure of LaN calculated with HSE and spin-orbit coupling. The conduction band and the valence band are separated by a direct gap of 0.62 eV at X. The valence band of LaN is mainly derived from N 2p orbitals and the conduction band is from the spatially extended unoccupied La 5d orbitals, as can be seen from the orbital projected band structure in Figure S2. Spin-orbit coupling has a profound effect on the structure of the valence band, particularly at the high symmetry points ($\Gamma$, X, and W). This arises from the contribution of La p orbitals near those points (Figure S2). After including spin-orbit effect, the band gap decreases from 0.75 eV to 0.62 eV. The 3-fold degenerate N p orbitals at $\Gamma$ split into 2-fold degenerate $p_{3/2}$ band and 1-fold $p_{1/2}$ band. Similarly, the 2-fold degeneracy between the second and third valence band at the X point and between the first and second valence bands at the W point is broken, with energy splittings of 340 meV and 301 meV, respectively. Table 2 summarizes the structural and electronic properties of LaN. We obtain a lattice constant of 5.282 Å, which is in excellent agreement with the experimental value of 5.284 Å.[30] The conduction band of LaN is dispersive with a bandwidth of several eV, leading to small electron masses especially along the transverse X-W and X-U directions. The hole effective masses are comparable to or even lighter than the hole effective masses of GaN.[31] The energy splitting between the two topmost valence bands is 85 meV, which is larger than the thermal energy $k_B T$ at room temperature of 26 meV. This indicates a much weaker phonon-mediated hole scattering from the valence band maximum to the second-highest valence band, as in the case of strained BAs.[32] The reduced scattering rate coupled with light effective masses may lead to a high hole mobility in LaN, although no p-type conductivity has been observed experimentally so far.

TABLE 2. Structural and electronic properties of LaN evaluated with the HSE06 hybrid functional and spin-orbit coupling.

| Crystal Structure | | Lattice constant (Å) | |
|---|---|---|---|
| Rocksalt | | 5.282 | |
| Band gap (eV) | | Valence band splitting (meV) | |
| 0.62 | | 85 | |
| Static dielectric constant, $\varepsilon_0$ | | High frequency dielectric constant, $\varepsilon_\infty$ | |
| 13.32 | | 9.04 | |
| Electron and hole effective masses | | | |
| | X-G | X-W | X-U |
| $m_e^*/m_0$ | 1.211 | 0.178 | 0.136 |
| $m_{h1}^*/m_0$ | 0.329 | 0.632 | 0.376 |
| $m^*_{h2}/m_0$ | 1.956 | 0.246 | 0.149 |
| $m^*_{h3}/m_0$ | 0.720 | 0.399 | 0.254 |

To investigate the band offsets between LaN and the other III-N materials in heterostructures and to understand its properties for photocatalysis, we calculate its absolute band positions by aligning the bulk average electrostatic potentials to the vacuum region. Figure 2 shows the absolute band alignment for LaN in comparison to other III-N materials. The band gap and absolute band positions of LaN are similar to ScN, another nitride material with the rocksalt crystal structure, forming a type-II alignment at the interface with a relative band offset of 0.3 eV. Interestingly, the electron affinity is almost identical for LaN and GaN. This can find potential applications in LaN/GaN heterostructures where electrons can move across the interface without a potential barrier.

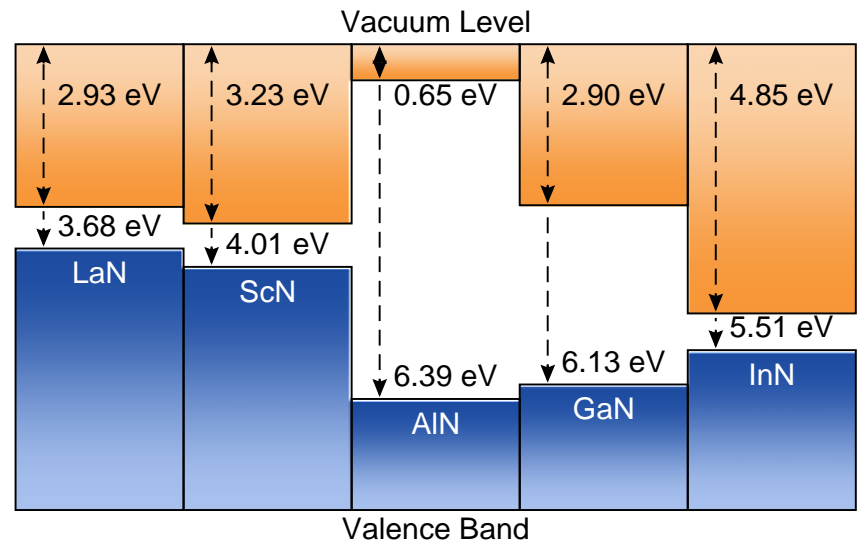

FIG. 2. Band alignment between LaN and other nitride materials. LaN has a type-II alignment with ScN and type-I alignment with GaN and AlN. The electron affinity of LaN and GaN is almost

identical. The band-alignment data of ScN and III-N are taken from Kumagai *et al*[33] and Moses *et al*,[34] respectively.

We then turn to the origin of the electrical conduction in LaN that is observed in experiment.[35–37] Previous studies attributed the conduction to the semi-metallic character of LaN. However, this explanation is inconsistent with our accurate band-structure results that find LaN to be a semiconductor with a band gap. An alternative explanation is that the electrical conduction in LaN is caused by unintentional doping either by intrinsic defects or by unintentional impurities incorporated during growth. To shed light on this issue, we calculate the formation energy of intrinsic defects (lanthanum vacancies $V_{La}$, nitrogen vacancies $V_N$) and common unintentional impurities (substitutional O and H interstitials). Figure 3 shows their formation energy as a function of the Fermi energy (referenced to the VBM) and growth conditions (N-rich or N-poor). Our key finding is that the donor-like defects ($V_N$ and $O_N$) have significantly lower formation energies (even negative) than acceptor-like defects, which identify LaN to be an intrinsically n-type semiconductor. Nitrogen vacancies act as negative-U double donors and are stable in the +2 charge state for Fermi levels throughout the majority of the band gap. O impurities are singly charged shallow donors, with an ionization energy of 50 meV. Therefore, both of these donor-like defects are candidate origins of the measured conductivity.

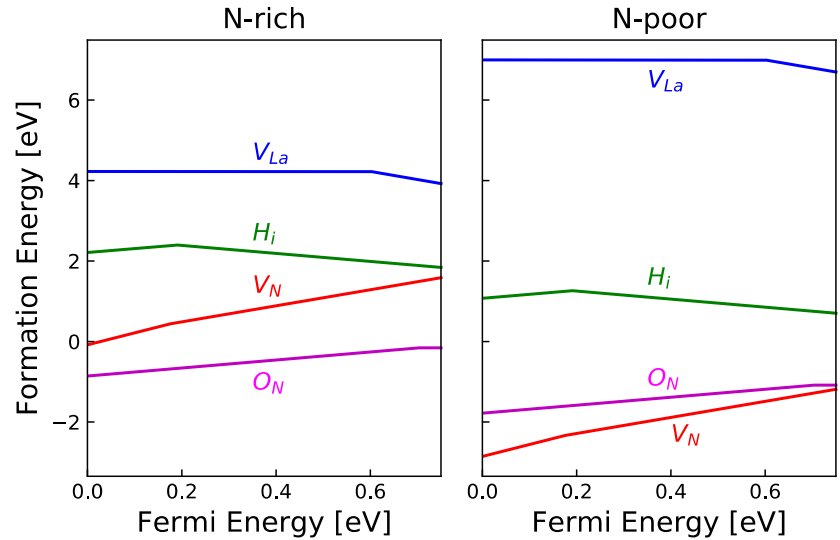

FIG. 3. Calculated defect formation energy as a function of Fermi level for the Nitrogen vacancy ($V_N$), Lanthanum vacancy ($V_{La}$), Hydrogen interstitial ($H_i$), and N-substitutional Oxygen impurity ($O_N$) in LaN. Donor-like defects ($V_N$ and $O_N$) exhibit the lowest formation energy for every Fermi level, indicating that LaN is an intrinsic n-type semiconductor and its conductivity originates from $V_N$ or $O_N$.

To further confirm the origin of the conductivity, we compare our findings to experimental measurements of the electrical conductivity as a function of temperature measured by four-point probe measurements in literature. Lesunova *et al* [36] find an exponentially increasing electrical conductivity with increasing temperature, which is characteristic of thermal activation of dopants in semiconductors. The temperature trend is also in contrast with the typical behavior in metals, in which the conductivity decreases with temperature due to increased carrier scattering by phonons. We further extract the donor activation energy by fitting the Arrhenius equation

$$\sigma(T) = \sigma_0 \exp\left(-\frac{E_A}{k_B T}\right) \qquad (2)$$

to the experimental electrical-conductivity measurements by Lesunova *et al*,[36] shown in Figure 4. The fitted value for the activation energy $E_A$ is 40 meV, which is in excellent agreement with our calculated ionization energy for shallow donors (39 meV) evaluated with the Bohr model

$$E_A = \frac{13.6\text{eV}\, m_e^*}{\varepsilon_r^2} \qquad (3)$$

using a directionally averaged electron effective mass of 0.51 $m_e$ and the static dielectric constant of 13.32. Based on this qualitative and quantitative agreement with the literature-reported experimental conductivity measurements, we conclude that the electrical conductivity of LaN originates from unintentional doping, and that $V_N$ or $O_N$ are likely candidate defects.

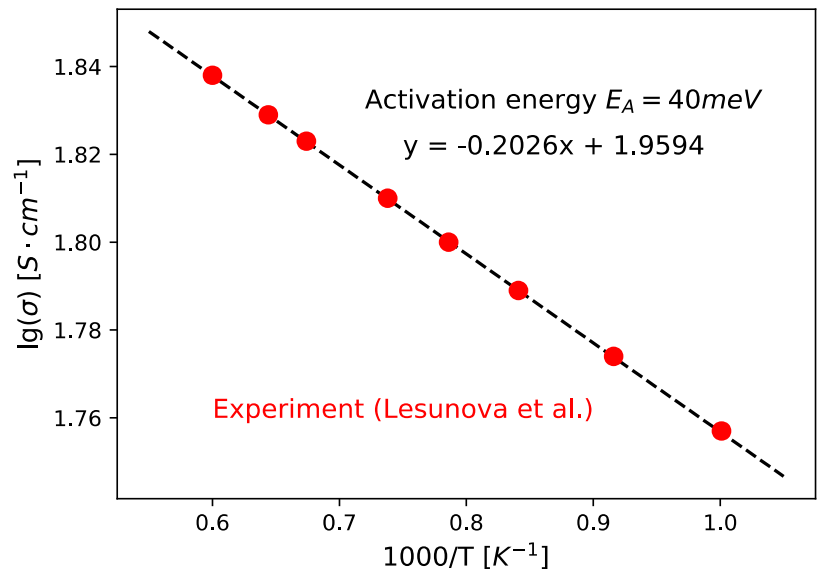

FIG. 4. Electrical conductivity of LaN as a function of temperature. The experimental data are taken from the measurements of Lesunova *et al*. [36]. The dashed line is a least-square fit to the conductivity data. The donor activation energy extracted from the fitting is 40 meV, in good agreement with the ionization energy for $O_N$ and $V_N$.

In our calculation, $O_N$ has a negative formation energy under both N-rich and N-poor conditions. However, this should not be interpreted as a sign of instability for LaN. In fact, the synthesis chemistry of LaN is well understood, and LaN has been utilized in catalysts for $NH_3$ synthesis[4] and for electrodes in supercapacitors.[37] Rather, the negative formation energy indicates that substitutional O is a major source of unintentional impurities which could possibly lead to degenerate n-type doping for LaN. In comparison, rocksalt ScN and Sc-containing nitride alloys, compounds with similar chemistry as LaN, have been found to have negative formation energy for $O_N$ in defect calculations,[33] and O gets unintentionally incorporated during growth[38,39] and results in degenerate n-type doping.[40] Thus, an oxygen-free growth environment is necessary to prevent the undesired degenerate doping by substitutional O for LaN.

In conclusion, we study the electronic properties of LaN using DFT calculations based on the HSE06 hybrid functional. In contrast to the semi-metallic nature claimed by most previous studies, we find that LaN is a direct-band-gap semiconductor with a gap of 0.62 eV at the X point. The light electron and hole effective masses and the near-zero conduction-band offset with GaN make LaN promising for electronic and optoelectronic applications. Our defect calculations indicate that LaN is intrinsically n-type and the source of the measured electrical conductivity is attributed to unintentional doping by nitrogen vacancies or substitutional oxygen. Our studies clarify the semiconducting nature of LaN, and reveal candidate unintentional donors that explain the origin of its measured electrical conductivity.

**Supplementary Material**

See Supplementary Material for the convergence of the $O_N$ defect formation energy and charge-transition level as a function of simulation cell size, the calculated band structure with the PBE functional, and the orbital-projected HSE band structure.

**Acknowledgements**

This study was supported by the National Science Foundation through Grant No. DMR-1561008. The band-structure and band-alignment calculations used resources of the National Energy Research Scientific Computing (NERSC) Center, a Department of Energy Office of Science User Facility supported under Contract No. DEAC0205CH11231. The defect calculations used Comet and Data Oasis at the San Diego Supercomputer Center (SDSC) through allocation TG-DMR200031, an Extreme Science and Engineering Discovery Environment (XSEDE)[41] user facility supported by National Science Foundation grant number ACI-1548562.
## DATA AVAILABILITY

The data that support the findings of this study are available from the corresponding author upon reasonable request.

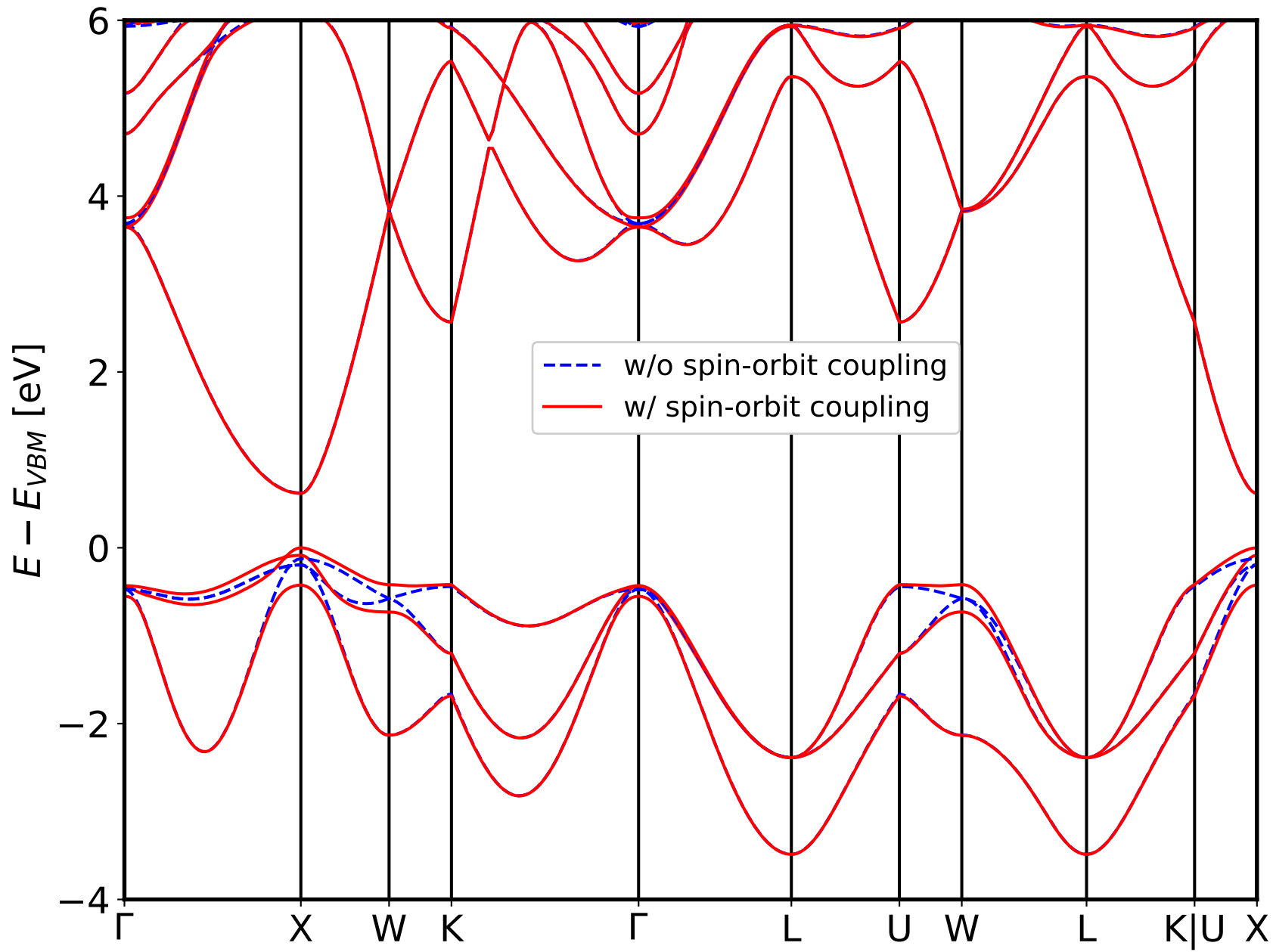

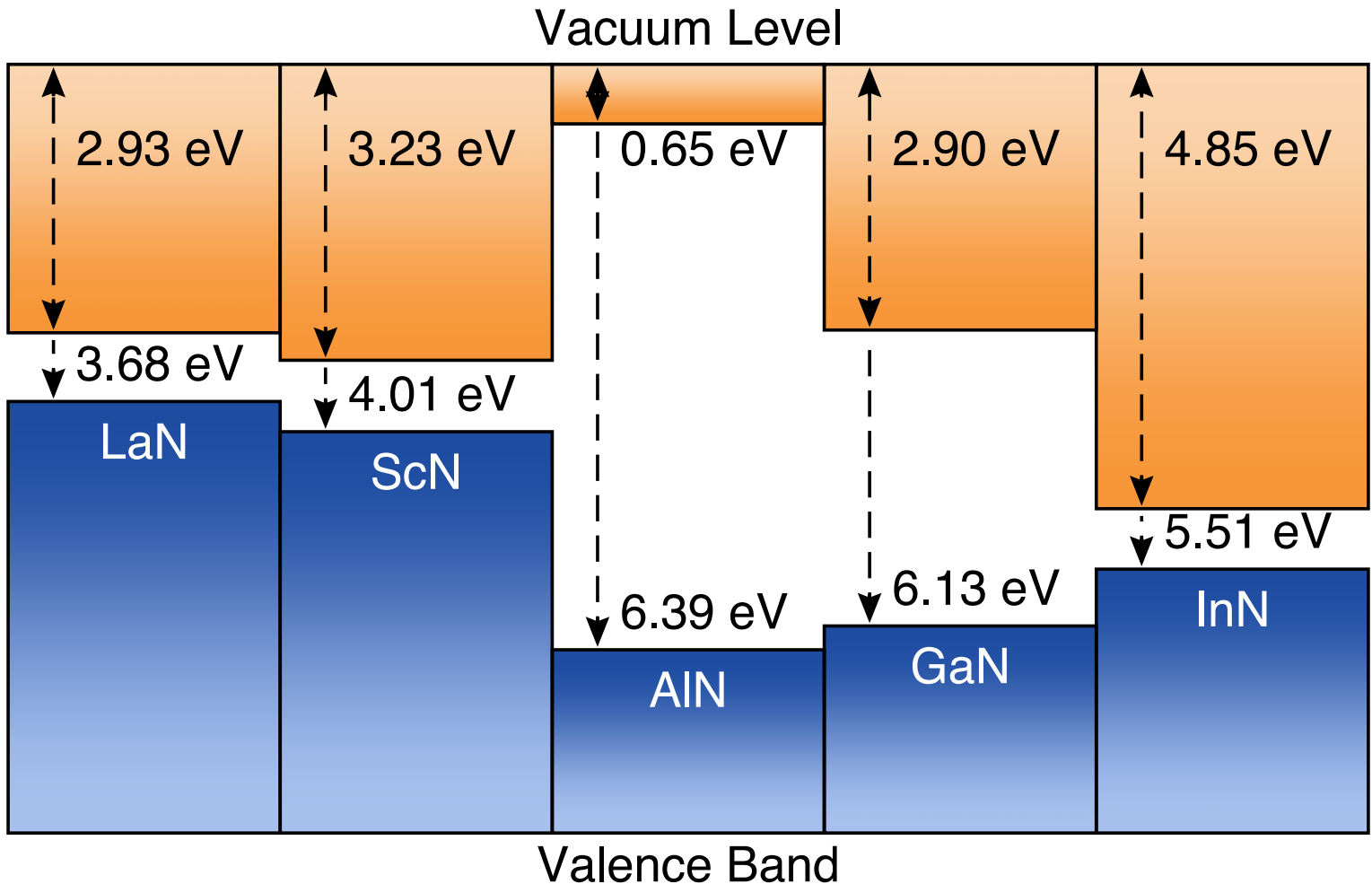

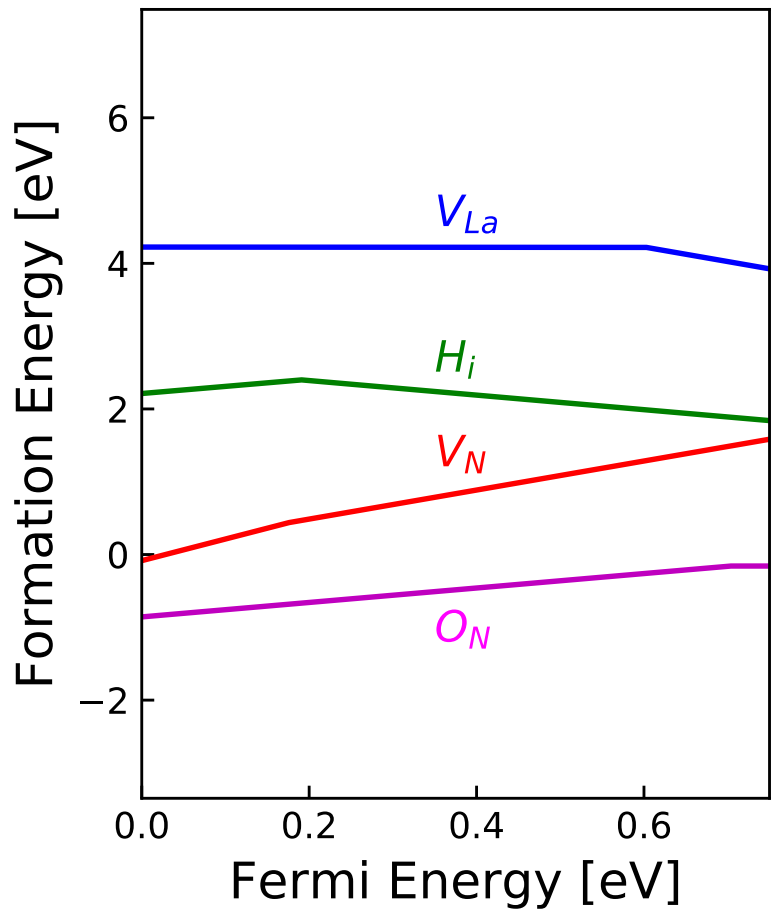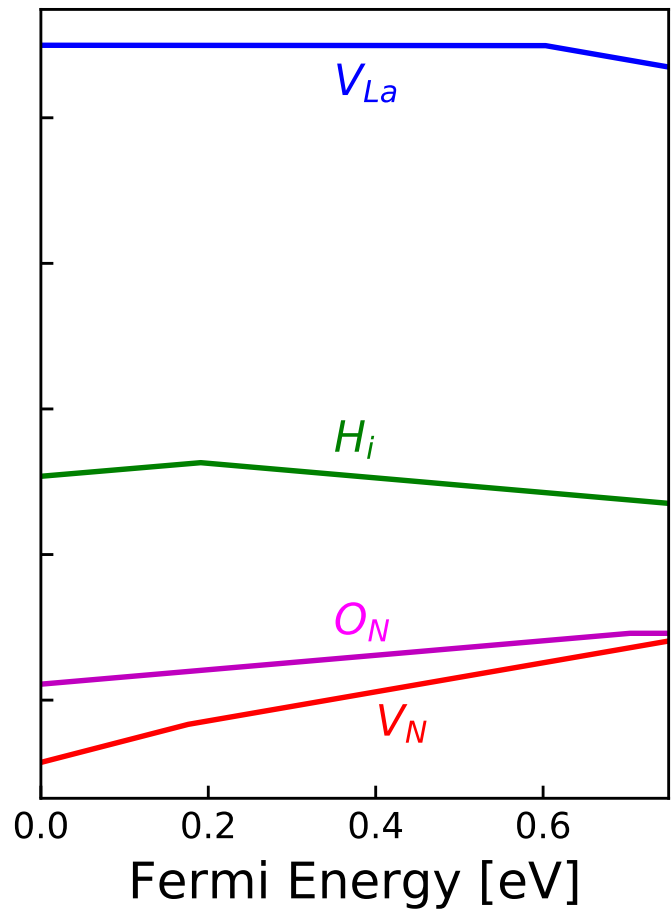

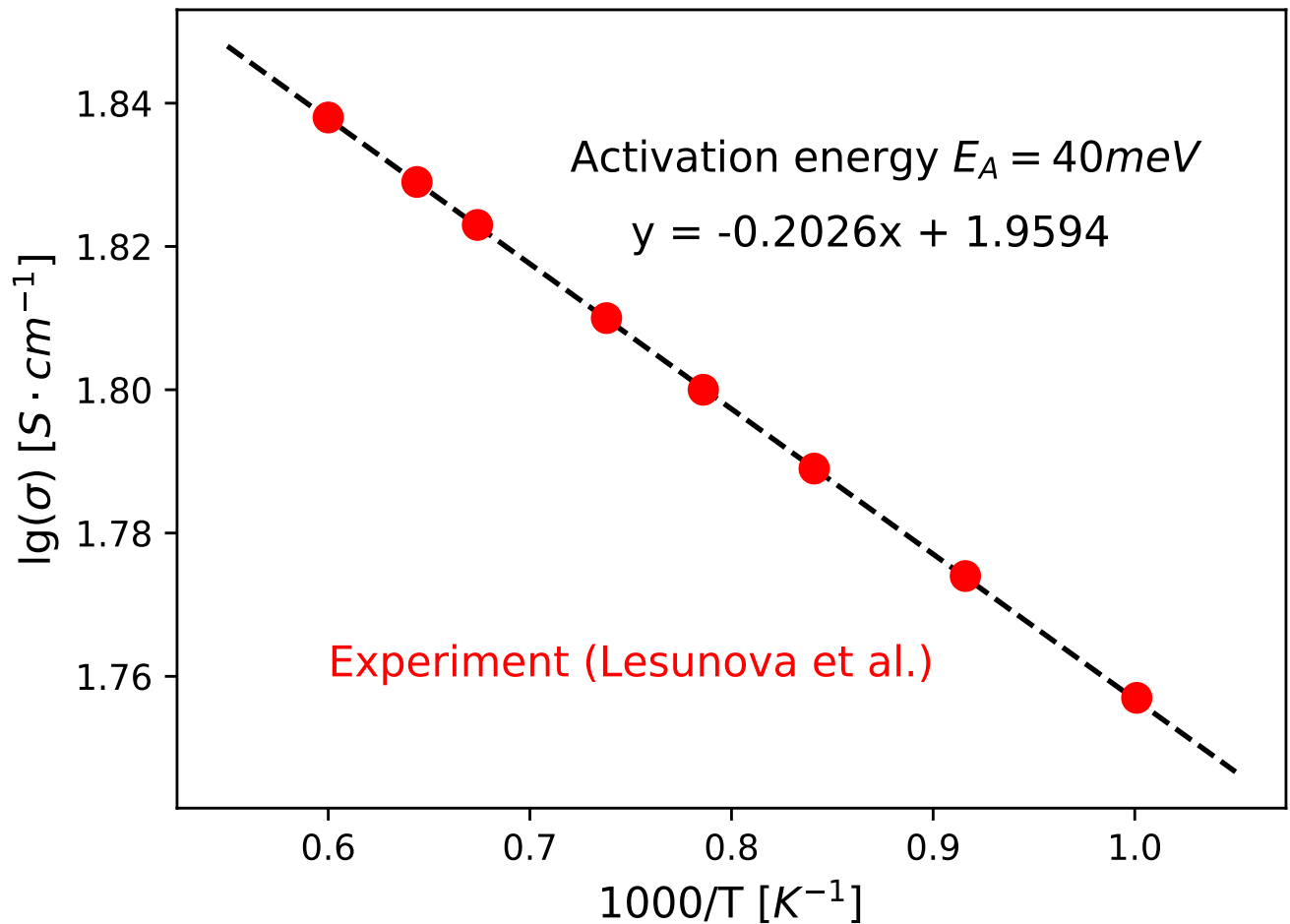